\documentclass[aps,twocolumn,amsmath,amssymb,reprint,numbers,superscriptaddress,noeprint]{revtex4-1}

\usepackage{soul}
\usepackage{geometry}
\usepackage{graphicx}
\usepackage{booktabs}
\usepackage{array}
\usepackage{textcase}
\usepackage{relsize}
\usepackage{color}
\usepackage{xcolor}
\usepackage{eqnarray}
\usepackage{caption}
\usepackage{textcomp}
\usepackage{setspace}
\usepackage{subfig}
\usepackage{tikz}
\usetikzlibrary{positioning}
\usepackage{hyperref}
\hypersetup{
     colorlinks   = true,
     linkcolor    = blue,
     citecolor    = blue,
     urlcolor = blue
}

\definecolor{cola}{rgb}{0.7,0.1,0.1}
\definecolor{colb}{rgb}{0.9,0.4,0}
\definecolor{colc}{rgb}{0.3,0.7,0}
\definecolor{cold}{rgb}{0,0.35,0.75}
\definecolor{cole}{rgb}{0.63, 0.13, 0.94}

\begin{document}

\title[]{Coherence properties of shallow donor qubits in ZnO}

\author{Xiayu Linpeng}
\email{lpxy1992@uw.edu}
\affiliation{Department of Physics, University of Washington, Seattle, Washington 98195, USA}
\author{Maria L. K. Viitaniemi}
\affiliation{Department of Physics, University of Washington, Seattle, Washington 98195, USA}
\author{Aswin Vishnuradhan}
\affiliation{Department of Applied Physics and Quantum-Phase Electronics Center (QPEC), University of Tokyo, Tokyo 113-8656, Japan}
\author{Y. Kozuka}
\affiliation{Department of Applied Physics and Quantum-Phase Electronics Center (QPEC), University of Tokyo, Tokyo 113-8656, Japan}
\affiliation{JST, PRESTO, Kawaguchi, Saitama 332-0012, Japan}
\author{Cameron Johnson}
\affiliation{Department of Physics, University of Oregon, Eugene, Oregon 97403, USA}
\author{M. Kawasaki}
\affiliation{Department of Applied Physics and Quantum-Phase Electronics Center (QPEC), University of Tokyo, Tokyo 113-8656, Japan}
\author{Kai-Mei C. Fu}
\email{kaimeifu@uw.edu}
\affiliation{Department of Physics, University of Washington, Seattle, Washington 98195, USA}
\affiliation{Department of Electrical Engineering, University of Washington, Seattle, Washington 98195, USA}

\begin{abstract}
Defects in crystals are leading candidates for photon-based quantum technologies, but progress in developing practical devices critically depends on improving defect optical and spin properties. Motivated by this need, we study a new defect qubit candidate, the shallow donor in ZnO. We demonstrate all-optical control of the electron spin state of the donor qubits and measure the spin coherence properties. We find a longitudinal relaxation time T$_1$ exceeding 100 ms, an inhomogeneous dephasing time T$_2^*$ of $17\pm2$~ns, and a Hahn spin-echo time T$_2$ of $50\pm13$~\textmu s. The magnitude of T$_2^*$ is consistent with the inhomogeneity of the nuclear hyperfine field in natural ZnO.  Possible mechanisms limiting T$_2$ include instantaneous diffusion and nuclear spin diffusion (spectral diffusion). These results are comparable to the phosphorous donor system in natural silicon, suggesting that with isotope and chemical purification long qubit coherence times can be obtained for donor spins in a direct band gap semiconductor. This work motivates further research on high-purity material growth, quantum device fabrication, and high-fidelity control of the donor:ZnO system for quantum technologies.

\end{abstract}

\date{\today}

\maketitle

\section{Introduction}
Defect centers in crystals have attracted significant attention as qubit candidates for quantum communication~\cite{ref:heshami2016qme,ref:alibart2016rai} and computation~\cite{ref:ladd2010qc} due to the ability to realize spin-photon entanglement and scalable device integration. A two-node network, the fundamental building block for measurement-based quantum computation~\cite{ref:benjamin2009pmb,ref:nickerson2013tqc,ref:raussendorf2001owq} and long-range quantum communication~\cite{ref:sangouard2011qrb,ref:briegel1998qrr}, can be generated via a single photon measurement on two non-interacting, spatially separated qubits. Measurement-based entanglement generation requires qubits with high optical efficiency, the ability for single qubit control, and long spin coherence times. The negatively charged nitrogen-vacancy (NV) center~\cite{ref:santori2010nqo,ref:doherty2013nvc_x} is one of the leading candidates for these protocols and two-node networks have been demonstrated, albeit with low entanglement generation rates~\cite{ref:bernien2013heb}. Factors limiting the achievable rates are optical inhomogeneity, spectral diffusion, and low zero-phonon radiative efficiency. Whilst numerous efforts are focused on overcoming these challenges in the NV system~\cite{ref:schroder2016qnd}, searching for new defect centers with better properties is an alternative solution. Donors in isotope purified $^{28}$Si have shown promising features such as ultra-long coherence times~\cite{ref:saeedi2013rtq,ref:watson2017aee} and high fidelity qubit control~\cite{ref:muhonen2015qqg}. However, the indirect band gap of Si makes photon-mediated entanglement and therefore the development of scalable quantum networks challenging~\cite{ref:benjamin2009pmb,ref:nickerson2014fsq,ref:tosi2017sqp}. Studies of donors in direct band-gap III-V materials have shown efficient optical transitions and demonstrated spin control and readout~\cite{ref:fu2006msf,ref:fu2008ucd,clark2009uos}, but their electron spin coherence times are limited by hyperfine interactions with the host nuclear spins~\cite{ref:fu2005cpt} and spin-orbit coupling~\cite{Linpeng2016}. Donors in direct band-gap II-VI semiconductors similarly boast efficient optical transitions~\cite{ref:wagner2011bez} and, as we show here in ZnO, can exhibit long coherence times. Critical for long-term qubit viability is the compatibility of ZnO with microfabrication processing~\cite{ref:Ozgur2005crz,ref:Djuriic2010zno} and the possibility of entanglement generation between the ZnO donor electron and donor/lattice nuclei based on the hyperfine interaction~\cite{ref:gonzalez1982mrs}. This electron-nucleus register, demonstrated in both P:Si~\cite{ref:steger2012qis} and NV:diamond systems~\cite{robledo2011hfp}, enables deterministic network scaling in the presence of large photon loss~\cite{ref:benjamin2009pmb,ref:benjamin2006bgs}.

In this paper, we measure the relaxation and coherence properties of an ensemble of Ga donors in ZnO. Ensemble spin initialization is demonstrated using resonant continuous-wave (cw) excitation. The longitudinal spin relaxation time T$_1$ shows a B$^{-3.5}$ relationship, dominated by a spin-orbit mediated phonon interaction. The longest T$_1$ observed in the experiment is $\sim$0.1~s at 2.25~T, with T$_1$ increasing with decreasing field. Coherent spin control of donor electrons is achieved with ultra fast optical pulses, red-detuned from the neutral donor (D$^0$) to donor-bound exciton (D$^0$X) resonance. The D$^0$ coherence is then probed via all-optical Ramsey interferometry and spin-echo measurements~\cite{clark2009uos}. The inhomogeneous dephasing time T$_2^*$ is measured to be 17 $\pm$ 2~ns which is consistent with the theoretical estimates of inhomogeneous electron-nuclear hyperfine interaction in natural ZnO. The effect of the inhomogeneous nuclear field is suppressed by a spin echo sequence with a measured spin-echo time T$_2$ of $50\pm13$~\textmu s at 5~T. Possible mechanisms limiting T$_2$ include spectral diffusion due to flip-flops of $^{67}$Zn nuclear spin pairs~\cite{desousa2003tni} and instantaneous diffusion due to the rephasing pulse in the spin echo sequence~\cite{ref:tyryshkin2011esc}.  

\begin{figure*}[!htbp]
  \centering
  \includegraphics[width=6.5in]{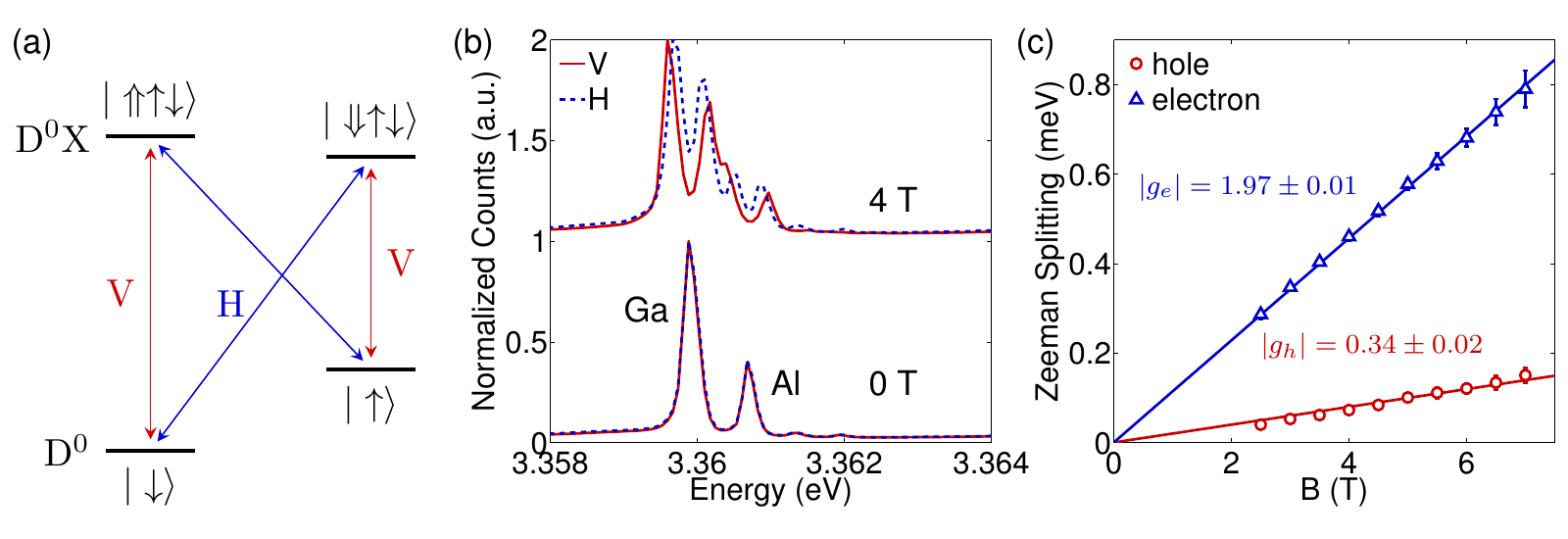}
  \caption{\label{spectra_fig}  (a) Energy diagram of the donor system at magnetic field in the Voigt geometry. V and H represent vertical polarization ($\hat{\varepsilon} \perp \vec{B}$) and horizontal polarization ($\hat{\varepsilon} \parallel \vec{B}$), respectively. $|\Uparrow\rangle (|\uparrow\rangle)$ denotes the hole (electron) spin. (b) Spectra at 0~T and 4~T with V and H polarized collection. The excitation laser is at 3.446~eV with vertical polarization. Temperature is 5.5 K. Both the Ga and Al donor peaks split into 4 different peaks with applied magnetic field. (c) Electron and hole Zeeman splitting of the Ga donor as function of magnetic fields. The red and blue lines are linear fits of the Zeeman splitting. For these data, both the excitation and collection spot sizes are $\sim$1~\textmu m. 
  }
\end{figure*}

\section{Setup and Photoluminescence spectrum}
The ZnO sample studied in this paper is a 360~\textmu m thick Tokyo Denpa ZnO crystal. The sample included a 0.7~\textmu m high-purity ZnO epilayer grown by molecular beam epitaxy~\cite{ref:akasaka2010mfl}, however the measurement signal was dominated by substrate donor emission. The total donor concentration is on the order $10^{17}$~cm$^{-3}$, determined by capacitance-voltage measurements~\cite{ref:nakano2007sco}. The sample is mounted in a continuous flow cryostat with a superconducting magnet in Voigt geometry, i.e. $\hat{c} \perp \vec B$, where $\hat{c}$ is the optical propagation axis. $\hat{c}$ is parallel to the [0001] direction of the ZnO crystal. All measurements are performed at temperatures between 1.5 and 5.5~K. 

The energy diagram of the shallow donor in a magnetic field is shown in Fig.~\ref{spectra_fig}(a). The D$^0$ spin states split due to the electron Zeeman effect. The Zeeman splitting of the D$^0$X state is solely determined by the hole spin, as the two bound electrons form a spin singlet. Typical spectra at 0~T and 4~T are shown in Fig.~\ref{spectra_fig}(b). At 0~T, the two main peaks  correspond to Al donors (3.3607~eV) and Ga donors (3.3599~eV)~\cite{Meyer2004}. To further confirm the two peaks are from donors, PL spectra with resonant excitation are taken to demonstrate the correlation between the main donor peaks and the corresponding two electron satellite transitions~\cite{supplementary2018}, i.e. transitions from the D$^0$X to the $2s$ and $2p$ D$^0$ orbital states. At 4~T, the Al and Ga peaks each split into 4 peaks due to the electron and hole Zeeman splitting. The polarization dependence of the 4 peaks confirms the $\Gamma_7$ valence band symmetry assignment~\cite{Wagner2009}. The measured $g$-factors for the Ga donors are $|g_e| = 1.97\pm0.01$ and $|g_h| = 0.34\pm0.02$, determined by linear fits of the electron and hole Zeeman splitting at different fields, as shown in Fig~\ref{spectra_fig}(c). For the remainder of the paper, we will focus on the Ga donor. 

The $\Gamma_7$ D$^0$X transitions, which exhibit short radiative lifetimes $\sim$1~ns and $>$90\% efficiency in zero-phonon emission~\cite{ref:wagner2011bez}, provides a natural $\Lambda$ system for Raman-based photon-heralded entanglement schemes~\cite{ref:cabrillo1999ces}. The ability to utilize other valence-band D$^0$X transitions~\cite{Meyer2004} to realize highly desirable cycling transitions and ``L''-shaped systems will be investigated in future work.  


\begin{figure}[!htbp ]
  \centering
  \includegraphics[width=3.5in]{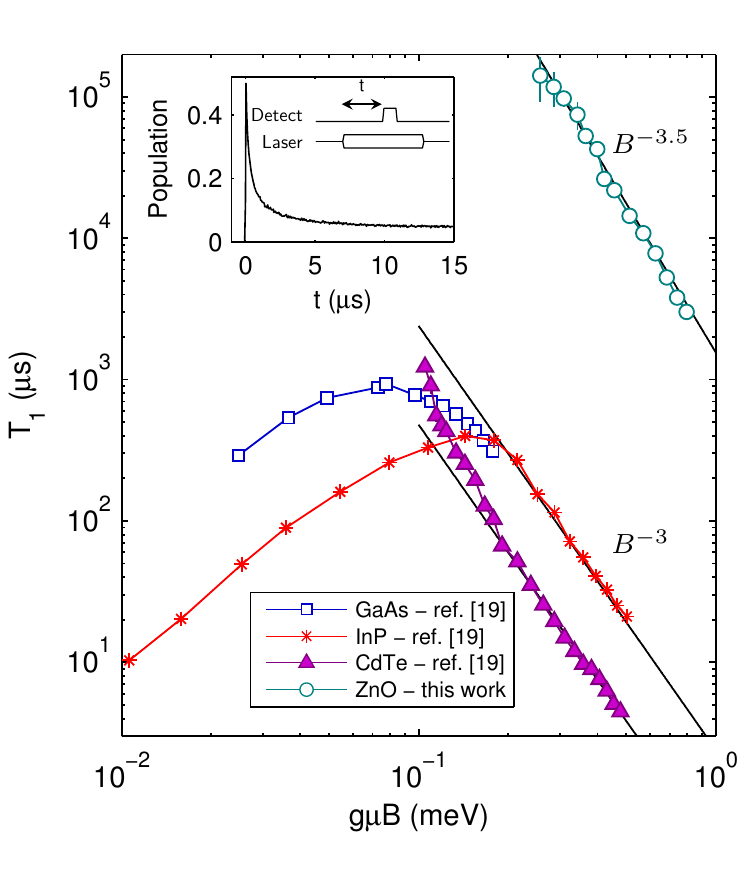}
  \caption{\label{T1fig} The longitudinal spin relaxation time T$_1$ as a function of the Zeeman energy for donors in GaAs, InP, CdTe and ZnO. The temperature is at 1.5~K. The data for GaAs, InP and CdTe is reproduced from a prior work~\cite{Linpeng2016}. The inset shows a typical ZnO optical pumping curve at 5~T and the corresponding laser sequence. The PL is detected by an avalanche photodiode with a 50~ns timing resolution. For the ZnO data, both the excitation and collection spot sizes are $\sim$1~\textmu m.}
\end{figure}

\section{Spin initialization and T\texorpdfstring{$_1$}{TEXT} measurement}
Spin initialization, the first step to utilize the spin as a qubit, is performed by optical pumping. In our experiment, a 10~\textmu s cw pulse is resonantly applied on the transition $|\uparrow\rangle \leftrightarrow |\Downarrow\uparrow\downarrow\rangle$ to initialize the electron spin state to $|\downarrow\rangle$. To visualize the optical pumping, the spins are first prepared with equal population in $|\uparrow\rangle$ and $|\downarrow\rangle$ using a scrambling pulse, i.e. a high power laser pulse with photon energy higher than the donor transitions. A typical optical pumping curve is shown in the inset of Fig.~\ref{T1fig}. An estimate of the pumping efficiency using the contrast ratio of the optical pumping curve~\cite{robledo2011hfp} yields a fidelity of 95\% at 1.5~K and 5~T. The efficiency of the optical pumping decreases with decreasing magnetic field. At low field, the Zeeman energy becomes comparable to the optical linewidth of the D$^0$X transitions. In this case, population in $|\downarrow\rangle$ can be simultaneously pumped back to $|\uparrow\rangle$, decreasing the optical pumping efficiency. For this reason, we are only able to observe an optical pumping signal at fields larger than 2.25~T. 

\begin{figure*}[t]
  \centering
  \includegraphics[width=6.5in]{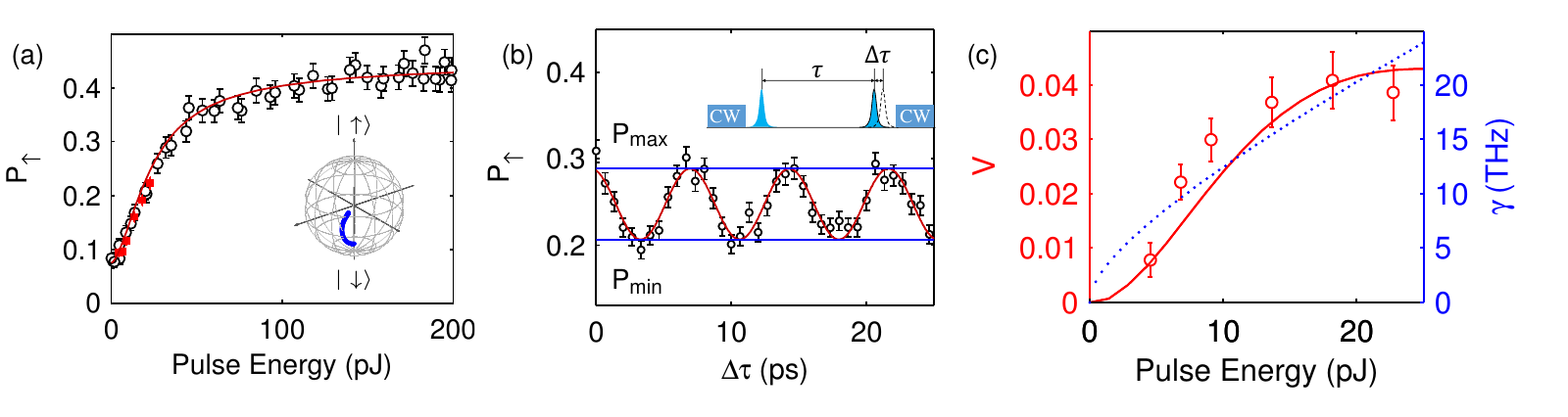}
  \caption{\label{rabifig} (a) P$_{\uparrow}$ (population of $|\uparrow\rangle$) as a function of the single-pulse energy with spin initialized to $|\downarrow\rangle$ and then excited by a 1.9~ps pulse. Data points represented by red squares at low powers are taken at the same power as data points in {\bf c}. The red curve is a simultaneous least squares fit for data in {\bf a} and {\bf c}. The inset shows how the state changes in the Bloch sphere using the simulated results. (b) A typical Ramsey interference pattern with 18~pJ pulse energy. The inset shows the laser sequence, where $\tau$ is the delay between the two pulses ($\tau=0.8$~ns in this data). The first cw pulse initializes the spin and the second cw pulse is to used to read out. (c) The Ramsey fringe amplitude V~=~(P$_{\text{max}}$~-~P$_{\text{min}}$)/2 as a function of the single pulse energy. The red line is the simulation result from the simultaneous fit. The blue dotted line shows the fit parameter $\gamma$ (excited state dephasing rate) as a function of pulse energy. For these data, the excitation spot size is $\sim$2~\textmu m, the collection spot size is $\sim$0.6~\textmu m. The temperature is at 1.5 K and the magnetic field is at 5 T. The 1.9~ps ultra-fast pulses are detuned by $\Delta /2\pi=$~3.57~THz from the transition $|\downarrow\rangle \Leftrightarrow |\Downarrow\uparrow\downarrow\rangle$.
  }
\end{figure*}

T$_1$ is measured by recording the population recovery to thermal equilibrium after spin initialization. T$_1$ at 1.5~K as function of magnetic field is shown in Fig.~\ref{T1fig}, with previous measurement results in GaAs, InP and CdTe~\cite{Linpeng2016} included for comparison. In the high-field region, the strong inverse power dependence on B indicates that relaxation is induced by phonon interactions, mediated by electron spin-orbit coupling~\cite{Khaetskii2001}. The high B-field dependence in ZnO is similar to what is observed in the other three semiconductors. However, T$_{1,\mathrm{ZnO}}$ is over two orders of magnitude longer as a result of lower spin-orbit coupling. At low field, a positive B-field dependence of T$_1$ is observed in GaAs and InP due to the short electron correlation time at the donor sites~\cite{Linpeng2016}. In ZnO, this mechanism is expected to be weaker because of the small electron Bohr radius. The high B-field dependence, together with the small Bohr radius, suggest T$_1$ can approach and possibly exceed seconds at lower magnetic fields. Control of the spin at lower fields will require a high-purity sample with narrow optical linewidth, as optical pumping can only be efficient if the linewidth is much smaller than the Zeeman splitting.



\section{Optical spin coherent control}
In the next series of measurements we use ultrafast optical pulses to create and probe the electron spin coherence. To obtain both strong optical pumping efficiency and long T$_1$, we choose an intermediate magnetic field to study, i.e. 5~T. At 5~T, the large electron Zeeman splitting (138 GHz) makes direct microwave control of the electron spin challenging. An alternative is to use a detuned ultra-fast optical pulse to coherently rotate the spins~\cite{Clark2007qcb}, which can be understood using a 4-level density matrix model. For the 4-level donor system, the Hamiltonian in the interaction picture with the rotating wave approximation is
\begin{eqnarray}
H = 
\begin{pmatrix} 
0 & 0 & -\frac{\Omega_{13}(t)}{2} & -\frac{\Omega_{14}(t)}{2} \\ 
0 & \omega_e & -\frac{\Omega_{23}(t)}{2} & -\frac{\Omega_{24}(t)}{2} \\
-\frac{\Omega_{13}^*(t)}{2} & -\frac{\Omega_{23}^*(t)}{2} & \Delta & 0 \\
-\frac{\Omega_{14}^*(t)}{2} & -\frac{\Omega_{24}^*(t)}{2} & 0 & \Delta+\omega_h
\end{pmatrix},
\label{eq:4lv}
\end{eqnarray}
where $\omega_e$($\omega_h$) is the energy of the electron (hole) Zeeman splitting, $\Delta$ is the red detuning between the ultra-fast laser and the transition $|\downarrow\rangle \Leftrightarrow |\Downarrow\uparrow\downarrow\rangle$, $\Omega_{ij}(t)=~\overrightarrow{\mu_{ij}}\cdot\overrightarrow{E}(t)/\hbar$ is the product of the electric field and the dipole matrix element of transition $|i\rangle \Leftrightarrow |j\rangle$ ($i$ = 1, 2, 3, 4 corresponding to states $|\downarrow\rangle$, $|\uparrow\rangle$, $|\Downarrow\uparrow\downarrow\rangle$, $|\Uparrow\uparrow\downarrow\rangle$). In the far-detuned limit ($\Delta \gg$ the optical pulse width), the populations of  the two excited states can be adiabatically eliminated~\cite{harris1994ric} and Eq.~\ref{eq:4lv} reduces to an effective 2-level Hamiltonian describing coherent rotations of the electron spin.

In our experiment, the polarization of the laser is adjusted so that $\Omega_{13}=\Omega_{23}=\Omega_{14}=\Omega_{24}=\Omega_R$~\cite{supplementary2018}. The ZnO donor effective Hamilitonian is then given by~\cite{supplementary2018}
\begin{equation}
H_{\mathrm{eff}} = \begin{pmatrix} 0 & \frac{\Omega_{\mathrm{eff}}(t)}{2} e^{-i \omega_e t} \\ \frac{\Omega_{\mathrm{eff}}^*(t)}{2} e^{i \omega_e t} & 0 \end{pmatrix},
\end{equation}
where $\Omega_{\mathrm{eff}} = \frac{|\Omega_R|^2}{2}(\frac{1}{\Delta} + \frac{1}{ \Delta+\omega_h})$ is the effective Rabi frequency. The axis of the rotation is determined by the timing of the pulse due to the $e^{\pm i\omega_e t}$ terms in $H_{\text{eff}}$. 
While this 2-level model provides intuition for how a single optical pulse coherently rotates the spin, it does not consider decoherence or relaxation. To analyze the dynamics of the density matrix in a more accurate way, we use the full 4-level master equation with decoherence and relaxation taken into consideration, i.e. $\partial \rho/\partial t = - i [H, \rho] + L(\rho)$, where $L(\rho)$ is the Lindblad operator~\cite{supplementary2018}. All data in Fig.~\ref{rabifig} is fit using this 4-level master equation model.

To generate a coherent superposition of the ground spin states, we first optically pump the donors to $|\downarrow\rangle$. A 1.9~ps pulse generated from a mode-locked Ti:Sapphire laser is frequency doubled to obtain the ultra-fast control pulse. Figure~\ref{rabifig}(a) shows the dependence of $|\uparrow\rangle$ population after the ultrafast pulse as a function of the pulse energy. We attribute the saturation of the population transfer at high pulse powers to laser-induced dephasing between the D$^0$X states and the D$^0$ states. 

Due to the laser-induced dephasing, coherent rotations are only expected at low pulse energy. The coherence of the small-angle rotation can be probed via Ramsey interferometry. Standard Ramsey experiments are done by measuring the spin population after two $\pi/2$ pulses with variable delay between them. An oscillation of the spin population as a function of the delay time can be observed due to the Larmor precession of the electron spin. 
Though only small-angle rotations are accessible in our system, they can also produce Ramsey interference, albeit with smaller oscillation amplitude.
A representative Ramsey fringe using small-angle rotations is shown in Fig.~\ref{rabifig}(b). The fit oscillation frequency in Fig.~\ref{rabifig}(b) is $136 \pm 3$~GHz at 5 T, which matches the predicted $137.9\pm0.7$~GHz using the measured electron g-factor. The Ramsey fringe amplitude as a function of the pulse energy is shown in Fig.~\ref{rabifig}(c). A least squares fit based on the 4-level density matrix model is used to fit the data in Fig.~\ref{rabifig}(a) and (c) simultaneously. 
The fit parameters are the ratio between the pulse energy and the peak of $\Omega_R(t)^2$, and the parameters $\beta_{1,2}$ which describe the laser-induced excited state dephasing $\gamma = \beta_1 \Omega_R(t) + \beta_2 \Omega_R^2 (t)$~\cite{supplementary2018}.
While the mechanism for this dephasing is unknown, one possibility is the unintentional excitation of real carriers. The fit slightly underestimates the fringe amplitude in Fig.~\ref{rabifig}(c). We attribute it to the uneven pulse power across the collection spot, leading to an inhomogeneity in the spin rotation angle~\cite{supplementary2018}.

Ultra-fast optical spin-control is a powerful tool to probe the coherence of the electron spins and measure the coherence time, however long-term it will be necessary to achieve high fidelity full-angle control for quantum applications. A possible solution is to utilize spin-resonant microwave fields, which has been successfully demonstrated in NV centers and donors in Si. For practical devices in ZnO, we must decrease our magnetic field such that the electron Zeeman splitting of the ground states is less than 10~GHz. This is difficult in our current sample due to the large inhomogeneous optical linewidth which makes optical pumping inefficient at lower magnetic fields. This challenge can be overcome with higher purity samples or single donor isolation.

	
\begin{figure}[!htbp ]
  \centering
  \includegraphics[width=3.5in]{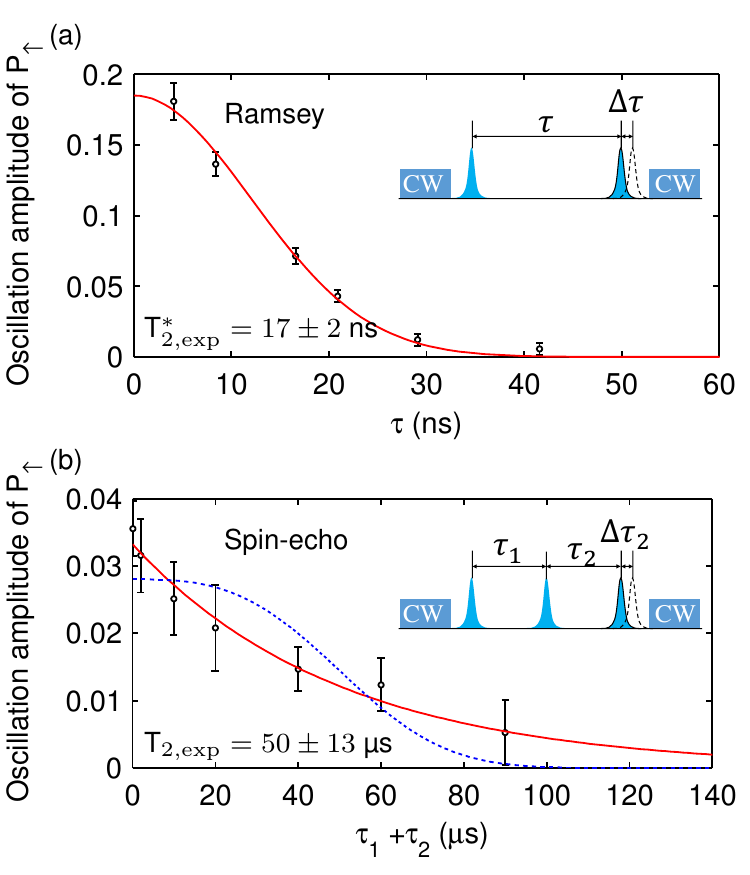}
  \caption{\label{T2fig}  (a) The Ramsey fringe amplitude, i.e. the oscillation amplitude of P$_{\uparrow}$ in the Ramsey interference pattern, is measured as a function of delay time $\tau$. The red curve shows a fit to $\exp(-(\tau/T_2^*)^2)$, giving T$_{2,\mathrm{exp}}^* = 17 \pm 2$~ns. (b) Spin-echo measurement of the dephasing time T$_2$. The delay $\tau_1~\simeq~\tau_2$. Oscillations of P$_{\uparrow}$ are observed by changing $\Delta \tau_2$. The oscillation amplitude of P$_{\uparrow}$ is measured as a function of $\tau_1+\tau_2$. The red curve shows a fit to $\exp(-\frac{\tau_1+\tau_2}{T_2})$, giving T$_{2,\mathrm{exp}} = 50 \pm 13$~\textmu s. For comparison, the blue dashed line shows a fit to $\exp(-(\frac{\tau_1+\tau_2}{T_2})^3)$, the expected form for spectral diffusion. For these data, both the excitation and collection spot sizes are $\sim$0.5~\textmu m. The temperature is at 5.5~K and the magnetic field is at 5~T.  }
\end{figure}

\section{\texorpdfstring{T$_2^*$ and T$_2$ measurement}{TEXT}}
T$_2^*$ is extracted from the decay of the Ramsey fringe amplitude as a function of the pulse delay time, as shown in Fig.~\ref{T2fig}(a). A fit using $\exp(-(\tau/\text{T}_2^*)^2)$ gives T$_{2, \mathrm{exp}}^* = 17\pm2$~ns. This dephasing time originates from the inhomogeneous nuclear field due to the hyperfine interaction between electrons and lattice nuclear spins. For the Ga donors in ZnO, this includes the hyperfine interaction from both the Ga nucleus and the $^{67}$Zn nuclei. T$_2^*$ can be estimated from the dispersion of the hyperfine field $\Delta_B $ with $\text{T}_2^* = \hbar/g_e \mu_B \Delta_B$~\cite{merkulov2002esr}. As only one Ga nucleus is in the effective wave function of the electron bound to the donor, the effective field from Ga has 4 different values due to 3/2 nuclear spin of Ga: 
\begin{equation}
B_{\mathrm{Ga}} = \frac{2\mu_0}{3 g_e}  \frac{\mu_{\mathrm{Ga}}}{I_{\mathrm{Ga}}}|u_{\mathrm{Zn}}|^2 |\psi(0)|^2 \times \{\frac{3}{2},\frac{1}{2},-\frac{1}{2},-\frac{3}{2}\}.
\label{eq:Bga}
\end{equation}
The hyperfine field due to the numerous $^{67}$Zn nuclei is estimated to have a Gaussian dispersion $\Delta_{B, \mathrm{Zn}}$~\cite{merkulov2002esr}:
\begin{equation}
\Delta_{B,\mathrm{Zn}} = \frac{\mu_0\mu_{\mathrm{Zn}}}{g_e}\sqrt{\frac{32}{27}}\sqrt{\frac{I_{\mathrm{Zn}}+1}{I_{\mathrm{Zn}}}}|u_{\mathrm{Zn}}|^2\sqrt{f\sum_j |\psi(\vec R_j)|^4},
\label{eq:Bzn}
\end{equation}
In Eqs.~\ref{eq:Bga} and \ref{eq:Bzn}, $\mu_B$ is the Bohr magneton, $g_e$ is the electron g-factor, $\mu_0$ is the vacuum permeability. $I_{\mathrm{Zn}} = 5/2$ ($I_{\mathrm{Ga}} = 3/2$) is the nuclear spin of $^{67}$Zn (Ga), $\mu_\mathrm{Zn} = 0.874 \mu_N$ ($\mu_\mathrm{Ga} = 2.24 \mu_N$) is the nuclear magnetic moment of $^{67}$Zn (Ga) and $\mu_N$ is the nuclear magneton. $f = 4.1$\% is the natural abundance of $^{67}$Zn. $\psi(\vec R_j)$ ($\psi(0)$) is the hydrogenic effective-mass envelope wave function of electron at the $j$th Zn (Ga) lattice site. $|u_{\mathrm{Zn}}|^2$ is the ratio of Bloch function density at the Zn site to the average Bloch function density. From electron spin resonance measurements in ZnO~\cite{ref:gonzalez1982mrs}, $|u_{\mathrm{Zn}}|^2\simeq1120$. Using the effective mass Bohr radius $a_\mathrm{B}\simeq1.7$~nm and by combining the hyperfine interactions from both Ga and $^{67}$Zn, we find $\text{T}_{2,\mathrm{theory}}^*\simeq 9$~ns~\cite{supplementary2018}, which is on the same order as our experimental result. Moving to isolated single donors in isotope-purified ZnO can eliminate this dephasing mechanism.


We next apply a spin echo sequence to suppress the effect of the inhomogeneous nuclear field. A standard spin echo includes two $\pi/2$ pulses separated by one $\pi$ pulse. 
It has be shown that three small angle rotations have a similar effect but with a smaller echo signal~\cite{clark2009uos}.
The measured spin-echo decoherence time is T$_{2,\mathrm{exp}}$ = $50\pm13$~\textmu s using an exponential fit, as shown in Fig.~\ref{T2fig}(b). Possible mechanisms limiting T$_2$ are instantaneous diffusion and spectral diffusion.

Instantaneous diffusion (ID) is the decoherence caused by the refocusing pulse in the spin-echo sequence. During the refocusing pulse, the dipole-coupled electron spins bound to different donors all rotate with the same angle. Therefore, the energy of this dipole-dipole interaction doesn't flip sign after the refocusing pulse and the phase cannot be corrected. The decay of the signal follows an $\exp(-t/\text{T}_{2,\mathrm{ID}})$ with T$_{2,\mathrm{ID}}$ given by~\cite{kurshev1992eos,tribollet2008uls}
\begin{equation}
1/\text{T}_{2,\mathrm{ID}} = \frac{\mu_0 (g_e \mu_B)^2 N_\text{Ga}}{9 \sqrt{3} \pi \hbar} \sin^2\frac{\theta_2}{2}
\end{equation}
where $N_\text{Ga}$ is the density of Ga donors and $\theta_2$ is the rotation angle of the refocusing pulse. Due to the comparable excitation and collection spot sizes in the experiment, the rotation angle varies across the collection spot making an accurate estimation of $\theta_2$ challenging. A reasonable range of $\theta_2$ is $\pi/5$~$\sim$~$\pi/2$. While the Ga donor concentration is uncertain, a chemical analysis of similar samples indicates a Ga donor density below 1~ppm. Using $N_{\mathrm{Ga}} \simeq 10^{16}$~cm$^{-3}$, T$_{2,\mathrm{ID}}$ ranges from 240~\textmu s to 1.27~ms. This is an underestimation as the refocusing pulse also affects the spin states of other donors and shallow impurities.



Spectral diffusion (SD) of the electron spin energy can occur due to flip-flops of dipole-coupled $^{67}$Zn nuclear spins. The measured T$_{2,\mathrm{exp}}^{\mathrm{ZnO}}$ is of similar magnitude to T$_2$ measured for phosphorous donors in natural Si~\cite{tyryshkin2006cos,witzel2006qtf}, which is limited by this spectral diffusion mechanism. Considering the similar isotope composition between ZnO and Si, we expect spectral diffusion to also be significant in ZnO. We estimate T$_{2,\mathrm{SD}}$ with a stochastic model developed for phosphorous donors in Si~\cite{chiba1972ese}. Assuming a Gaussian diffusion kernel, the decay of the signal exhibits an $\exp(-(t/\text{T}_{2,\mathrm{SD}})^3)$ dependence with T$_{2,\mathrm{SD}}$ given by
\begin{equation}
1/\text{T}_{2,\mathrm{SD}} \simeq \left[\frac{8 \pi}{27 \sqrt{3} \hbar} \mu_0 \mu_{\mathrm{Zn}} g_e \mu_B n \Sigma_j b^2_{j}\right]^{1/3},
\label{eq:Tsd}
\end{equation}
\begin{equation}
\Sigma_j b^2_{j} = f \frac{\mu_0^2}{16 \pi^2}\frac{\mu^4_{\mathrm{Zn}} }{\hbar^2} \Sigma_j \frac{(1-3 \cos^2\theta_j)^2}{r_j^6},
\end{equation}
where $n$ is the density of $^{67}$Zn. For a given $^{67}$Zn nucleus, $b_j$ is the dipole-dipole interaction between it and the $j$th $^{67}$Zn. $r_j$ is the distance between the two nuclei and $\theta_j$ is the angle between ${\vec r}_j$ and the B-field. Using Eq.~\ref{eq:Tsd}, we estimate T$_{2,\mathrm{SD}} \simeq$~200~\textmu s. 



The magnitude of T$_2$ estimated by both mechanisms is in reasonable agreement with T$_{2,\mathrm{exp}}$. While we find better agreement in the experimental decay shape with the instantaneous diffusion mechanism, as shown in Fig.~\ref{T2fig}(b), it is still hard to confirm the dominant mechanism considering the low signal-to-noise ratio and due to the fact that only one measurement of T$_{2}$ has been carried out. To rigorously determine the mechanism, future experiments measuring the dependence of T$_2$ on different parameters will be conducted, including the abundance of $^{67}$Zn~\cite{whitaker2010hpc}, the donor density~\cite{ref:tyryshkin2011esc}, the rotation angle of the rephasing pulse~\cite{tribollet2008uls} and the magnetic field direction~\cite{tyryshkin2006cos}. 
The determination of the mechanism is important as this can be generalized to other II-VI materials, thus aiding in the search for superior defect-based qubit candidates. Regardless of which mechanism dominates T$_{2}$ in ZnO, practical devices will require both isotope purification and lower donor densities.

\section{Outlook}
In summary, we demonstrate optical spin control and read-out of Ga donor qubits in a bulk ZnO crystal. Long spin relaxation times (100~ms) and coherence times (50~\textmu s) are observed. These promising results motivate future work on the challenges toward making a practical quantum network out of optically-active donor qubits. In the ZnO donor platform, these challenges include chemical and isotope purification of the sample, high fidelity microwave control of the spin state, and single donor isolation. Thin films grown by molecular beam epitaxy have shown orders of magnitude lower impurity concentration than commercial ZnO substrates~\cite{ref:akasaka2010mfl}. Devices incorporating such high-purity layers will be essential for addressing all three challenges. In the near-term, single donor isolation for fundamental studies can be achieved in nanostructures fabricated by focused ion beam milling~\cite{ref:wu2004upc} or utilizing single nanowires~\cite{ref:johnson2003oce}. In the long term, scalable device integration  will require pushing ZnO fabrication techniques beyond the standard micro-fabrication techniques currently developed for ZnO~\cite{ref:Djuriic2010zno, ref:Pearton2004rap}.

\section{Acknowledgements}
This material is based upon work supported by the National Science Foundation under Grant No. 1150647, and in part upon work supported by the State of Washington through the University of Washington Clean Energy Institute and via funding from the Washington Research Foundation. K.-M. C. Fu acknowledges the support from the Research Corporation for Science Advancement as a Cottrell Scholar. Y. K. acknowledges JST, PRESTO Grant Number JPMJPR1763, Japan. We acknowledge Kelsey Bates for the measurement of the laser pulse length. We acknowledge Atsushi Tsukazaki and Joseph Falson for useful discussions on ZnO growth, properties, and characterization. 




 \newcommand{\noopsort}[1]{} \newcommand{\bibstar}{\textsuperscript{*}}

\end{document}